\documentclass[aps,prl,twocolumn,superscriptaddress,amssymb,amsmath,nofootinbib]{revtex4}

\usepackage{color}
\usepackage{graphicx}
\usepackage{float}
\usepackage[usenames,dvipsnames]{xcolor}

\begin{document}

\preprint{}

\title{Highly enhanced contributions of heavy Higgs bosons and new leptons\\ to muon $g-2$ and prospects at future colliders}

\author{Radovan Dermisek}

\email[]{dermisek@indiana.edu}

\author{Keith Hermanek}

\email[]{khermane@iu.edu}

\affiliation{Physics Department, Indiana University, Bloomington, IN, 47405, USA}

\author{Navin McGinnis}

\email[]{nmcginnis@triumf.ca}

\affiliation{High Energy Physics Division, Argonne National Laboratory, Argonne, IL, 60439}


\date{April 9, 2021}

\begin{abstract}
We show that in a two Higgs doublet model type-II extended by vectorlike leptons the contributions from heavy neutral and charged Higgs bosons to the anomalous magnetic moment of the muon
simultaneously  feature chiral enhancement from masses of new leptons and $\tan^2 \beta$  enhancement from couplings of Higgs bosons. Assuming moderate values of new Yukawa couplings, not exceeding one, that can remain perturbative to very high energy scales,  the measured value of muon $g-2$ can be explained within one standard deviation even with 6.5 TeV leptons or 20 TeV Higgs bosons. 
Allowing new couplings near the perturbativity limit, these mass ranges extend to 45 TeV for leptons and 185 TeV for Higgs bosons. 
In spite of the high scale of new physics  this scenario can be completely probed at planned future colliders.

\end{abstract}

\pacs{}
\keywords{}

\maketitle

\noindent
{\it Introduction.}
The discrepancy between the measured value of the muon anomalous magnetic moment, $(g-2)_{\mu}$,  and the standard model (SM) prediction, one of the largest discrepancies among precision electroweak (EW) measurements, is of the same order as the contributions of $Z$ and $W$ bosons. Thus, the most straightforward new physics explanations assume additional contributions from new particles with masses near the EW scale that couple to the muon with EW interactions, for reviews see Refs.~\cite{Czarnecki:2001pv, Endo:2013bba, Freitas:2014pua, Lindner:2016bgg}.

New particles near the EW scale are however increasingly constrained by direct searches at the Large Hadron Collider (LHC). 
Explanations with heavier particles can avoid constraints from direct searches but
require certain enhancements. Well known examples include the enhancement by the ratio of vacuum expectation values of the two Higgs doublets, $\tan \beta$, in supersymmetric models~\cite{Moroi:1995yh} and  chiral enhancement in models with vectorlike leptons~\cite{Kannike:2011ng, Dermisek:2013gta}. Alternatively, one can consider large couplings, making a given model non-perturbative not very far above the EW scale, or stretching other model parameters beyond what is typically considered. Even with enhancements or large couplings, it is not expected that new physics able to explain the discrepancy would be significantly above 1 TeV~\cite{Czarnecki:2001pv, Endo:2013bba, Freitas:2014pua, Lindner:2016bgg}. 
On the other hand, explanations with very light new particles are constrained by many kinds of experiments and, in order to survive, the new particles are SM singlets with specific interactions, see for example Refs.~\cite{Chen:2015vqy, Davoudiasl:2018fbb, Liu:2018xkx, Bauer:2019gfk}.

In this letter we show that in a two Higgs doublet model  type-II (2HDM-II), which is one of the simplest and most studied extensions of the SM Higgs sector~\cite{Gunion:1989we}, with vectorlike leptons mixing with the muon, 
there are contributions to $(g-2)_{\mu}$ originating from heavy neutral CP-even ($H$), CP-odd ($A$) and charged Higgs ($H^\pm$) bosons and new leptons which, in addition to chiral enhancement from masses of new leptons, are enhanced by $\tan^2 \beta$ compared to contributions of $W$, $Z$ and $h$ bosons and new leptons. As a result of this enhancement, we show that the measured value of $(g-2)_{\mu}$ can be explained within one standard deviation even with 6.5 TeV leptons or 20 TeV Higgs bosons assuming moderate values of all new Yukawa couplings not exceeding one, which can remain perturbative to very high energy scales, possibly the grand unification scale
(depending on other details of the model). Allowing new couplings near the perturbativity limit, $\sqrt{4\pi}$, the ranges of new lepton and Higgs masses able to explain the measured value of $(g-2)_{\mu}$ extend to 45 TeV and 185 TeV.

Interestingly, we will see that the high end of the suggested spectrum, which is well beyond the reach of the LHC and even the hadron (hh) option of the Future Circular Collider (FCC), could be probed indirectly at  the  International Linear Collider (ILC)  or the lepton (ee) option of the FCC.
However, another important consequence of the $\tan^2 \beta$ enhancement is the possibility to explain $(g-2)_{\mu}$ with  lighter new leptons and Higgs bosons but with highly reduced corrections to muon EW and Yukawa couplings, even close to three orders of magnitude smaller than current constraints allow. This is beyond the precision reach of the ILC and FCC-ee and thus testing this explanation of $(g-2)_{\mu}$ requires both precision and energy reach of future colliders, and illustrates their complementarity.

\noindent
{\it Model.}
We consider an extension of a two Higgs doublet model by vectorlike pairs of new leptons: SU(2) doublets $L_{L,R}$ and SU(2) singlets $E_{L,R}$, where $L_L$ and $E_R$ have the same quantum numbers as SM leptons. We assume that leptons couple to the two Higgs doublets as in the type-II model, namely the charged singlets couple to $H_d$, which can be enforced by a $Z_2$ symmetry~\cite{Gunion:1989we}. For simplicity, we further assume that only the 2nd generation mixes with vectorlike leptons which can be enforced by individual lepton number conservation. Alternatively, if mixing with all three generations is present, it is assumed that relevant couplings are sufficiently small to satisfy constraints from flavor changing processes. 
The most general Lagrangian consistent with our assumptions contains the following Yukawa and mass terms for the 2nd generation and vectorlike leptons:
\begin{flalign}
\mathcal{L}\supset& - y_{\mu}\bar{l}_L\mu_{R}H_{d} - \lambda_{E}\bar{l}_{L}E_{R}H_{d}  - \lambda_{L}\bar{L}_{L}\mu_{R}H_{d}   \nonumber \\   
& - \lambda\bar{L}_{L}E_{R}H_{d}   - \bar{\lambda}H_{d}^{\dagger}\bar{E}_{L}L_{R} \nonumber \\ 
	& - M_{L}\bar{L}_{L}L_{R} - M_{E}\bar{E}_{L}E_{R}  + h.c.,
\label{eq:lagrangian}	
\end{flalign}
where doublet components are labeled as: $l_{L}=( \nu_{\mu},  \mu_{L} )^T$,  $ L_{L,R}= ( L_{L,R}^{0}, L_{L,R}^{-})^T$, and $H_{d}= (H_{d}^{+}, H_{d}^{0})^T$. 
The neutral components of two Higgs doublets develop vacuum expectation values, $\left< H_u^0 \right> = v_u$ and $\left< H_d^0 \right> = v_d$, with $\sqrt{v_u^2 + v_d^2} = v = 174$ GeV and we define $\tan \beta \equiv v_u / v_d$. The charged lepton mass matrix becomes
\begin{equation}
	(\bar{\mu}_{L}, \bar{L}_{L}^{-}, \bar{E}_{L})\begin{pmatrix}y_{\mu}v_{d}&0 &\lambda_{E}v_{d}\\\lambda_{L}v_{d}&M_{L}&\lambda v_d\\0&\bar{\lambda}v_{d}& M_{E}\end{pmatrix}\begin{pmatrix}\mu_{R}\\ L_{R}^{-}\\ E_{R}\end{pmatrix}.
\end{equation}
This matrix can be diagonalized to obtain two new mass eigenstates, $e_4$ and $e_5$, in addition to the muon ($y_\mu$ is  determined iteratively to reproduce the muon mass for each choice of model parameters). We also label the new neutral mass eigenstate as $\nu_4$ (with mass given by $M_L$). As a result of mixing, couplings of the muon to $W$, $Z$, and $h$ are modified from their SM values, and couplings between the muon and heavy leptons are generated. General formulas can be found in \cite{Dermisek:2013gta,Dermisek:2015oja} and useful approximate formulas can be obtained from completely analogous expressions in the quark sector~\cite{Dermisek:2019vkc}.
Note that in the limit of heavy vectorlike leptons the lagrangian in Eq.~(\ref{eq:lagrangian}) reduces to 
\begin{equation}
 \mathcal{L}\supset - y_{\mu}\bar{l}_L\mu_{R}H_{d} - \frac{\lambda_L \bar \lambda \lambda_E}{M_L M_E}\bar{l}_L\mu_{R}H_{d}H_d^\dagger H_d + h.c.,
\end{equation}
where the higher dimensional operator is a new source of the muon mass that is directly related to the modification of $(g-2)_{\mu}$ and muon Yukawa coupling.

\begin{figure}[t]
\includegraphics[scale=0.4]{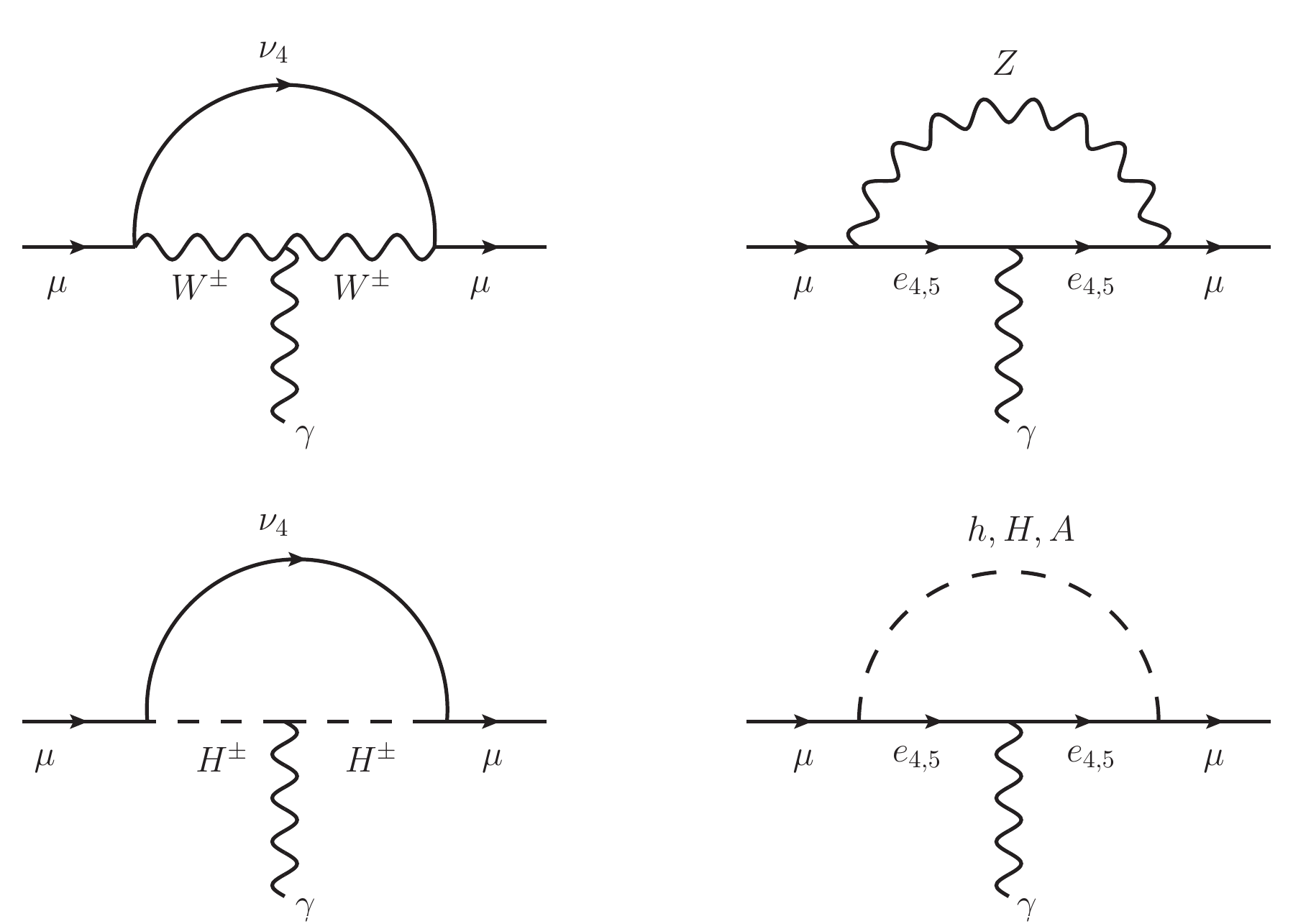}
\caption{$W$, $Z$, $h$ $H$, $A$, $H^\pm$ diagrams contributing to $(g-2)_{\mu}$.}
\label{fig:diags}
\end{figure}

\noindent
{\it Muon anomalous magnetic moment.}
With the recent measurement, the muon anomalous magnetic moment, $a_\mu \equiv (g-2)_{\mu}/2$,  sits at 4.2 standard deviations from the predicted value in the SM~\cite{Abi:2021gix, Aoyama:2020ynm},
\begin{equation}
\Delta a^{exp}_{\mu}\equiv a^{exp}_{\mu} - a_{\mu}^{SM} = (2.51 \pm 0.59)\times 10^{-9}.
\end{equation}
Mixing of the muon with new leptons results in additional contributions to $(g-2)_{\mu}$ from gauge and Higgs bosons and new leptons shown in Fig.~\ref{fig:diags}. Contributions from  $W$, $Z$ and $h$ loops were previously calculated in a one Higgs doublet model (1HDM) with the same matter content~\cite{Kannike:2011ng, Dermisek:2013gta}. For other examples of explanations of $(g-2)_{\mu}$ with vectorlike leptons, see also Refs.~\cite{Arnan:2016cpy, Megias:2017dzd, Kowalska:2017iqv, Raby:2017igl, Crivellin:2018qmi,  Kawamura:2019rth, Endo:2020tkb, Frank:2020smf, Chun:2020uzw}. Note that  2-loop Barr-Zee diagrams, recently discussed for example in Refs.~\cite{Frank:2020smf, Chun:2020uzw}, are negligible compared to the chirally and $\tan^2 \beta$ enhanced 1-loop contributions (they do not exceed $0.1\%$ of the total contribution for all presented scenarios and are typically much smaller than that). 

The complete calculation of  $(g-2)_{\mu}$ in the present model and other variations of 2HDM will be presented in Ref.~\cite{Dermisek:2021ajd}. The leading contributions of individual diagrams can be approximated by 
\begin{equation}
\Delta a^{i}_{\mu} 
\simeq   \frac{k^i}{16\pi^{2}}  \frac{m_\mu m_\mu^{LE}}{v^2} , \quad m_\mu^{LE} \equiv \frac{\lambda_{L} \bar{\lambda} \lambda_{E}}{M_{L}M_{E}} v^3 \cos^{3} \beta ,
\label{eq:dela}
\end{equation}
where $k^W =   1$, $k^Z =   -1/2$, $k^h = -3/2$, $k^H = - (11/12) \tan^2 \beta$, $k^A = - (5/12) \tan^2 \beta$, and $k^{H^\pm} = (1/3) \tan^2 \beta$. For $W$, $Z$ and $h$, these are very good approximations only for  $M_L \simeq M_E \gg m_Z$
~\cite{Dermisek:2013gta}. For heavy Higgs contributions, the approximations are very good for  $M_L \simeq M_E \simeq m_{H,A,H^\pm}$ up to  terms that cancel between $H$ and $A$ contributions when $m_{H} = m_A$. Note that $k^W + k^Z + k^h = -1$, while $k^H + k^A + k^{H^\pm} = - \tan^2 \beta$.

In a 1HDM, the $\cos^3 \beta$ factor is not present in Eq.~(\ref{eq:dela}) and thus it seems that a 2HDM-II can only generate a smaller contribution to $(g-2)_{\mu}$. However, when constraints from precision EW data are taken into account the advantage of the 2HDM becomes dramatic. Mixing of the muon with new leptons leads to modifications of muon couplings to $W$ and $Z$ bosons, known with $\sim 0.1\%$ precision~\cite{Zyla:2020zbs}, which translate  into 95\% C.L. bounds: $|\lambda_E v_d/ M_E  | \lesssim 0.03$ and $| \lambda_{L} v_d/M_L | \lesssim 0.04$~\cite{Kannike:2011ng, Dermisek:2015oja}, that enter $W$ and $Z$ couplings in squares. Both of these ratios  enter $m_\mu^{LE}$ in Eq.~(\ref{eq:dela}) and thus limit possible contributions to $(g-2)_{\mu}$.
In addition,  $m_\mu^{LE}$ directly affects the $h - \mu -\mu$ coupling, 
\begin{equation}
\lambda^h_{\mu \mu} \simeq y_\mu \cos \beta + 3 m_\mu^{LE}/v = ( m_\mu +2 m_\mu^{LE})/v ,
\label{eq:lambdah}
\end{equation}
that can significantly alter the SM prediction for $h \to \mu\mu$ rate. Thus, the experimental constraint on $\lambda^h_{\mu \mu} $~\cite{Aad:2020xfq} also limits possible contributions to $(g-2)_{\mu}$.  Note that $m_\mu^{LE}$ is the contribution to the muon mass from mixing with $L$ and $E$, $m_\mu \simeq y_\mu v_d + m_\mu^{LE}$. Numerically, the explanation of $\Delta a^{exp}_\mu$ by $W$, $Z$ and $h$ contributions, in both 1HDM and 2HDM-II, requires $m_\mu^{LE}  \simeq  - m_\mu$ and thus their contribution to $\Delta a_\mu$ is directly correlated with the modification of muon couplings to $W$, $Z$ and $h$. 


The contributions of heavy Higgs bosons in 2HDM-II to $(g-2)_{\mu}$ are enhanced by $\tan^2 \beta$ compared to contributions of $W$, $Z$, and $h$. This by itself is not an advantage compared to 1HDM because $\cos^3 \beta \tan^2 \beta$ is still smaller than one. 
However, these contributions are also enhanced by $\tan^2 \beta$ compared to modifications of the muon couplings to $W$, $Z$ and $h$. Numerically, 
the explanation of $\Delta a^{exp}$ by $H$, $A$ and $H^\pm$ contributions requires $m_\mu^{LE}  \tan^2 \beta \simeq  - m_\mu$.  Thus, increasing $\tan \beta$ increases the fraction of heavy Higgs contributions to $(g-2)_{\mu}$, decreases the required $m_\mu^{LE}$ and leads to correspondingly smaller modifications of muon Yukawa and gauge couplings. This $\tan^2 \beta$ enhancement opens the possibility to explain the measured value of $(g-2)_{\mu}$ with very heavy new leptons and new Higgs bosons while still satisfying constraints from precision EW data, or with lighter new leptons and Higgs bosons but highly reduced corrections to muon couplings.

\begin{figure}[t]
\includegraphics[scale=0.2]{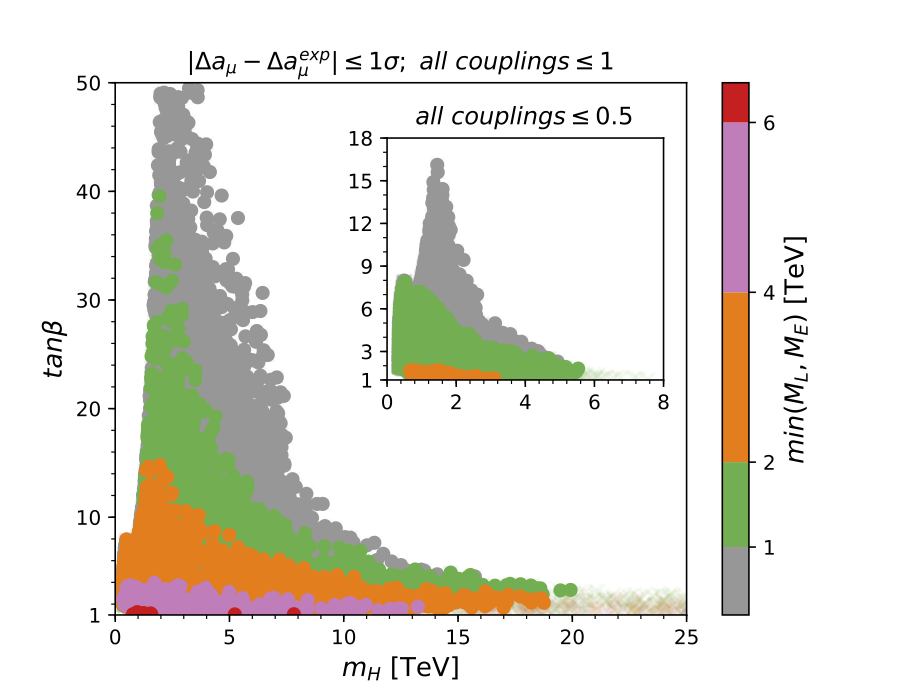} 
\includegraphics[scale=0.2]{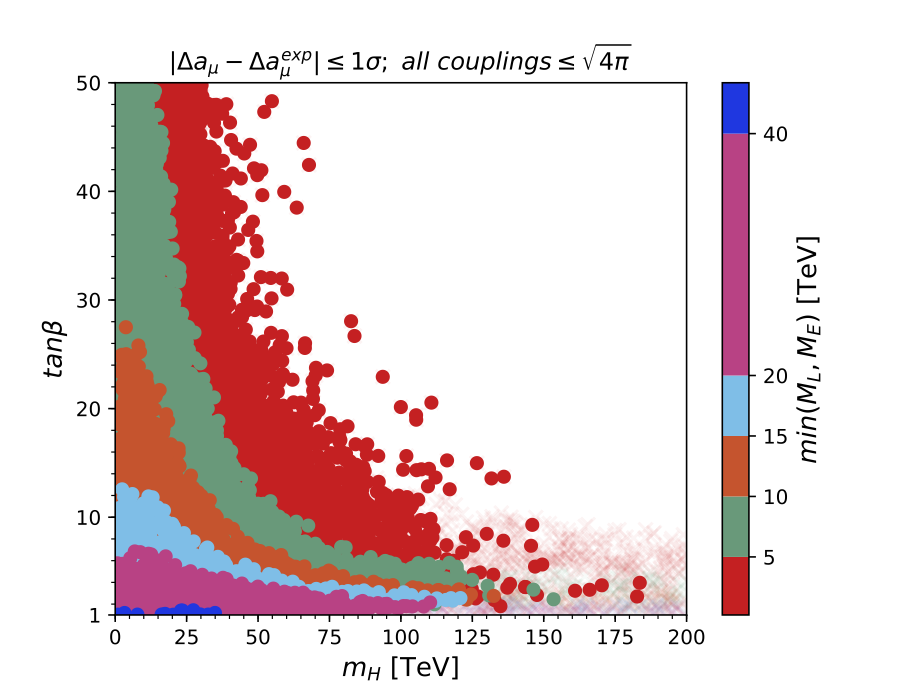}
\caption{Scenarios satisfying $\Delta a_{\mu}^{\exp}$ within $1\sigma$ in the $m_H - \tan \beta$ plane for Yukawa couplings not exceeding 1 (top), 0.5 (inset in top) and $\sqrt{4\pi}$ (bottom). Colors indicate masses of the lightest new lepton. Lightly shaded crosses without filled circles correspond to scenarios with heavy Higgs bosons contributing less than 50\% to $\Delta a_{\mu}$.}
\label{fig:MLE}
\end{figure}

To illustrate the first point, in Fig.~\ref{fig:MLE}  we plot the scenarios satisfying $\Delta a_{\mu}^{\exp}$ within $1\sigma$ in the $m_H - \tan \beta$ plane for  $\tan \beta \in~[1,50]$ with the size of any Yukawa coupling not exceeding 1 (top), 0.5 (inset in top) and $\sqrt{4\pi}$ (bottom). The upper limit on Yukawa couplings, $1$, is motivated by keeping them perturbative  to very large energy scales, while $\sqrt{4\pi}$ is the limit from perturbativity at the scale of new physics. 
In all numerical results we require $M_L > 800$ GeV, $M_E > 200$ GeV to generically satisfy constraints from direct searches for new leptons~\cite{Sirunyan:2019ofn, Aad:2020fzq, Sirunyan:2018mtv}.  However, it should be noted that the limits vary significantly with the assumed pattern of branching ratios of new leptons to $W$, $Z$ and $h$~\cite{Dermisek:2014qca} and, in the model we consider an arbitrary pattern of branching ratios can occur~\cite{Dermisek:2015hue}.
For simplicity we further assume $m_H = m_A = m_{H^\pm}$ and impose limits on  $H(A) \to \tau^+  \tau^-$~\cite{Aad:2020zxo} and  $H^+ \to t\bar b$~\cite{ATLAS:2020jqj} which are currently the strongest at large and small $\tan \beta$ respectively. With this assumption, the imposed limits are also sufficient to satisfy indirect constraints from flavor physics observables, see for example Ref.~\cite{Haller:2018nnx}.
In addition, we impose constraints from $h \to \mu\mu$~\cite{Aad:2020xfq} and precision EW data related to the muon: $Z$-pole observables, the $W$ partial width, the muon lifetime; and constraints from oblique corrections, summarized in Ref.~\cite{Zyla:2020zbs}. 
From the color coding in Fig.~\ref{fig:MLE} we see that assuming new Yukawa couplings not exceeding 1 ($\sqrt{4\pi}$) $\Delta a_{\mu}^{\exp}$  can be explained within $1\sigma$ even with 6.5 TeV (45 TeV) leptons or 20 TeV (185 TeV) Higgs bosons. 

\begin{figure}[t]
\includegraphics[scale=0.2]{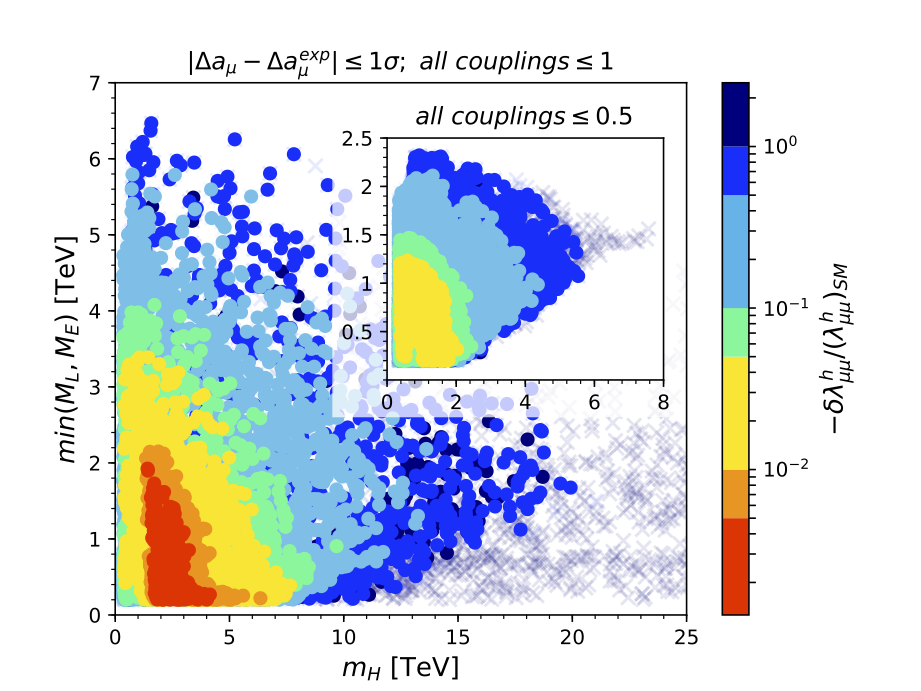}
\includegraphics[scale=0.2]{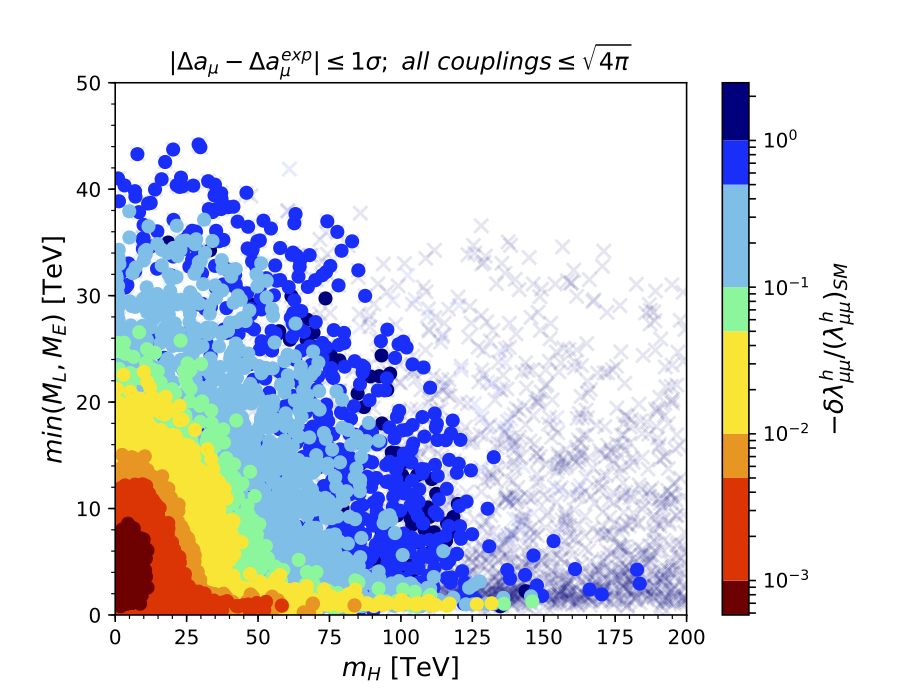}
\caption{The same scenarios as in Fig.~\ref{fig:MLE} in the $m_H - \min(M_L,M_E)$ plane with colors indicating $-\delta \lambda^h_{\mu \mu}/(\lambda^h_{\mu \mu})_{SM}$.}
\label{fig:hmumu}
\end{figure}

\begin{figure}[t]
\includegraphics[scale=0.2]{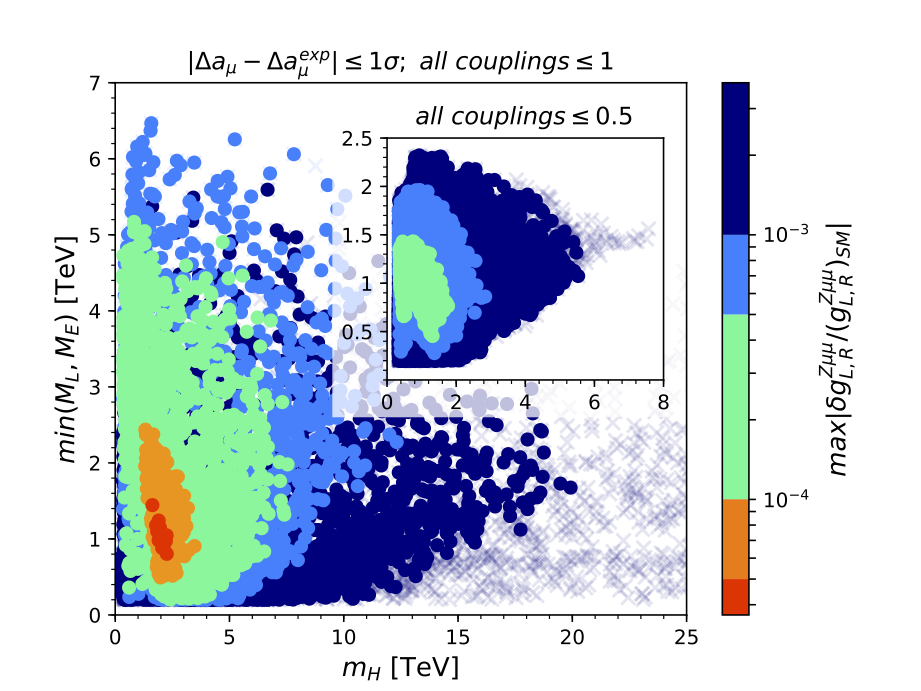}
\includegraphics[scale=0.2]{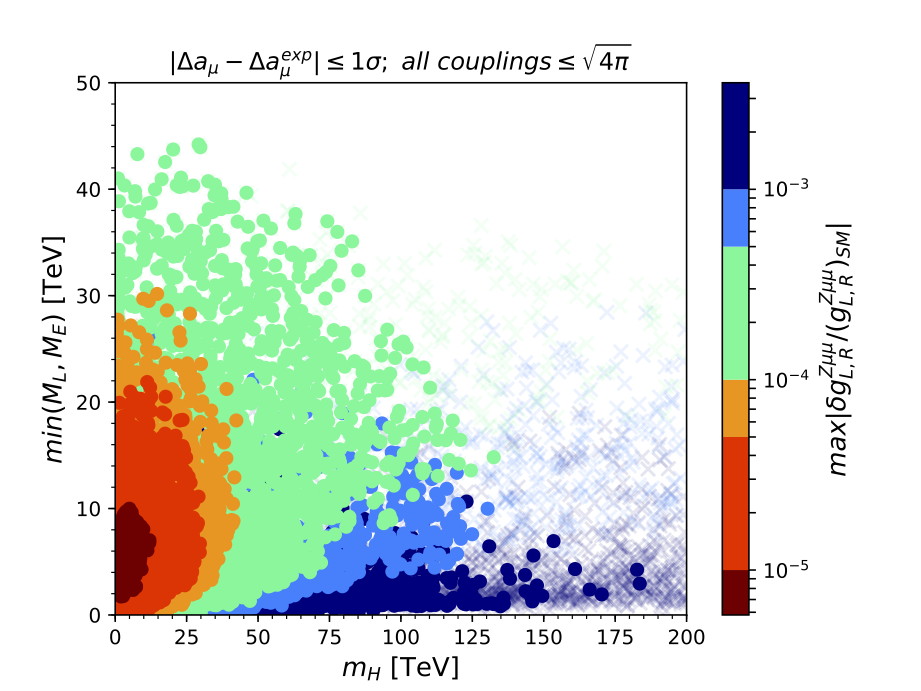}
\caption{The same scenarios as in Fig.~\ref{fig:MLE}  in the $m_H - \min(M_L,M_E)$ plane with colors indicating the largest relative modification of muon couplings to $Z$.}
\label{fig:ZWmu}
\end{figure}

To illustrate the second point, we plot the same scenarios in the $m_H - \min(M_L,M_E)$ plane with colors indicating $-\delta \lambda^h_{\mu \mu}/(\lambda^h_{\mu \mu})_{SM}$ in Fig.~\ref{fig:hmumu}  and the largest relative modification of muon couplings to the $Z$ boson in Fig.~\ref{fig:ZWmu}. Note that the model predicts negative correction to $\lambda^h_{\mu \mu}$ from the region of model parameters that can explain $\Delta a^{exp}_\mu$, and both $\delta \lambda^h_{\mu \mu} = 0 $ and $-2 (\lambda^h_{\mu \mu})_{SM}$ predict the same rate for $h \to \mu \mu$ as the SM. Thus, even large negative corrections are not currently excluded~\cite{Aad:2020xfq}. 

The heaviest spectrum explaining $\Delta a^{exp}_\mu$ corresponds to small $\tan \beta$ region, where the corrections to the muon Yukawa and gauge couplings are the largest, close to current bounds. As $ \tan \beta$ increases, in order to generate the same $\Delta a_\mu$ with the same size of new Yukawa couplings, the spectrum of new  leptons and Higgs bosons has to be lighter, and corrections to the muon Yukawa and gauge couplings decrease by $ \tan^2 \beta$. From Figs.~\ref{fig:hmumu} and~\ref{fig:ZWmu} we see that, assuming new Yukawa couplings not exceeding 1 or $\sqrt{4\pi}$, $\Delta a_{\mu}^{\exp}$ can be explained without affecting  $\lambda^h_{\mu \mu}$ at more than $10^{-3}$ or $5 \times 10^{-4}$ levels, and the muon couplings to the $Z$ boson at more than  $3 \times 10^{-5}$ or $5 \times 10^{-6}$ levels, respectively.

\noindent
{\it Discussion.} 
The possibility to explain $\Delta a^{exp}_\mu$ with very heavy new particles and moderate couplings indicates  that the constraints from direct searches and precision EW data can be generically satisfied for random choices of model parameters of specified order. The suggested spectrum of new leptons extending to tens of TeV and Higgs bosons close to 200 TeV is clearly well beyond the reach of the LHC and even 100 TeV FCC-hh. Even 6.5 TeV leptons (for new couplings up to 1) will likely not be seen at the FCC-hh~\cite{Bhattiprolu:2019vdu}. However, as a result of the discussed correlation between the scale of new physics and the modification of muon Yukawa and gauge couplings, the high end of the spectrum can be indirectly probed by precisely measuring these couplings. 


With 3 ab$^{-1}$ of data at the LHC, it is expected that $\lambda^h_{\mu \mu}$ will be measured with 5\% precision~\cite{Abada:2019lih} which  
will highly shrink the range of allowed masses, see Fig.~\ref{fig:hmumu}. However, even with the expected precision 0.5\% at the FCC-hh~\cite{Abada:2019lih}, scenarios with leptons and Higgs boson too heavy for direct detection would remain.

The muon couplings to the $Z$ boson could be measured with  0.05\% precision at 250 GeV ILC and  with  0.01\% precision with the GigaZ option~\cite{Fujii:2019zll}. 
 The reach of individual machines can be easily estimated from Fig.~\ref{fig:ZWmu}. For example, the GigaZ alone could completely cover all possibilities with new couplings up to 0.5. 
 Ultimately, at the FCC-ee, the precision for gauge couplings of the muon is expected to be at $\sim10^{-5}$ level~\cite{Abada:2019lih}. With this constraint alone only scenarios with 
 leptons and Higgs bosons within 11 TeV and 13 TeV, would remain viable. While such leptons are too heavy for the FCC-hh~\cite{Bhattiprolu:2019vdu}, the Higgs bosons in this range are likely within the reach (studies are needed). 
Although none of the proposed future colliders alone can fully probe this explanation of  $\Delta a^{exp}_\mu$, the combination of machines in precision and energy frontiers can.


It is also important to note, that similar effects, $\tan^2 \beta$ enhancements (at amplitude level) of loops of heavy Higgs bosons and new matter fields mixing with a SM fermion and the anti-correlation with the impact of the mixing on its SM gauge and Yukawa couplings, are expected for other observables in the charged lepton and down-type quark sectors. 


\vspace{0.05cm}

\acknowledgments
The work of RD was supported in part by the U.S. Department of Energy under grant number {DE}-SC0010120. NM was supported in part by the U.S. Department of Energy under contract No. DE-AC02-06CH11357.





\end{document}